\documentclass[a4paper,dvips]{jpconf}
\usepackage[dvips]{epsfig}
\usepackage{axodraw_c,color} 
\usepackage{graphics}
\usepackage{wrapfig}
\usepackage{floatfig}

\newcommand{\tfrac}[2]{{\textstyle{\frac{#1}{#2}}}}

\definecolor{green}{rgb}{0,0.7,0}
\definecolor{blue}{rgb}{0,0,1}
\definecolor{red}{rgb}{1,0,0}
\definecolor{brown}{rgb}{0.7,0.3,0}
\definecolor{violet}{rgb}{0.5,0,0.5}

\begin{document}
\title{Geometrical splitting and reduction of Feynman diagrams}

\author{Andrei I Davydychev$^{1,2}$}

\address{$^1$ Schlumberger, HFE, 
110 Schlumberger Drive, Sugar Land, Texas 77478, USA}
\address{$^2$ Institute for Nuclear Physics, Moscow State University,
119992 Moscow, Russia}

\ead{davyd@theory.sinp.msu.ru}

\begin{abstract}
A geometrical approach to the calculation of $N$-point Feynman diagrams is reviewed. 
It is shown that the geometrical splitting yields useful connections between 
Feynman integrals with different momenta and masses. It is demonstrated how these 
results can be used to reduce the number of variables in the occurring functions.
\end{abstract}

\vspace*{-130mm}
\begin{flushright}
MSU-SINP 2016-2/890\\
$\;$\\
\footnotesize{
Contribution to the Proceedings of ACAT-2016\\ 
(Valparaiso, Chile, January 18--22, 2016)
}
\end{flushright}
\vspace*{105mm}

\section{Introduction}

\begin{wrapfigure}{rb}{0.5\textwidth}
\vspace*{-5mm} 
  \begin{center}
  \vspace*{5mm}
  \includegraphics[width=18pc]{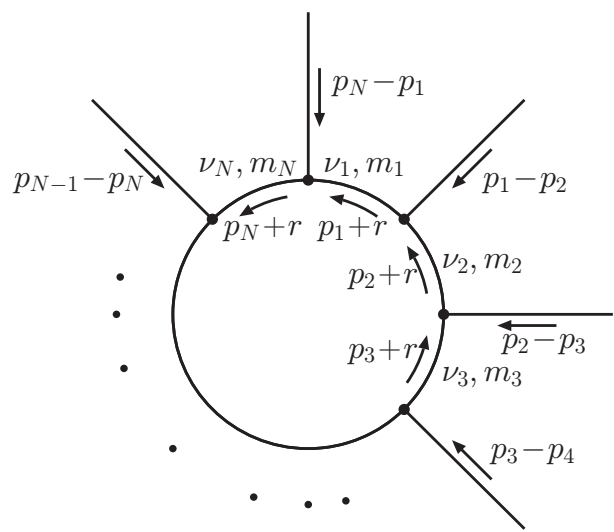}
  \vspace{8mm}
  \caption{\label{label}$N$-point diagram}
  \label{figNpt1}
  \end{center}
\end{wrapfigure}

A geometrical interpretation of kinematic invariants and other quantities
related to $N$-point Feynman diagrams (shown in figure~\ref{figNpt1}) helps us 
to understand the analytical structure of the results for these diagrams.
As an example, singularities of the general three-point function can be 
described pictorially through a tetrahedron constructed out of
the external momenta and internal masses. 
Such a geometrical visualization can be used to derive Landau equations defining
the positions of possible singularities \cite{Landau} 
(see also in \cite{3pt_sing}). 

In general the one-loop $N$-point diagrams (as shown in figure~\ref{figNpt1})
depend on $\tfrac{1}{2}N(N-1)$ momentum invariants $k_{jl}^2=(p_j-p_l)^2$
and $N$ masses of the internal particles $m_i$. Here and below we follow
the notation used in~\cite{JMP1-2}; in particular, the
powers of the internal scalar propagators are denoted as $\nu_i$, and 
the space-time dimension is denoted as $n$, so that we can also deal
with the dimensionally-regulated integrals with $n=4-2\varepsilon$ \cite{dimreg}.
Below we will mainly consider the cases when all $\nu_i=1$.  

In~\cite{DD-JMP,D-NIMA,Crete} it was demonstrated how such geometrical
ideas could be used for an analytical calculation of
one-loop $N$-point diagrams. For the geometrical interpretation, a ``basic simplex"
in $N$-dimensional Euclidean space is employed (a triangle for $N=2$, 
a tetrahedron for $N=3$, etc.), and the obtained results can be expressed
in terms of an integral over a $(N-1)$-dimensional spherical (or hyperbolic) 
simplex, which corresponds to the intersection of the basic simplex and the 
unit hypersphere (or the corresponding hyperbolic hypersurface), 
with a weight function depending on the angular distance $\theta$ between 
the integration point and the point 0, 
corresponding to the height of the basic simplex
(see in~\cite{DD-JMP}). 
For $n=N$ this weight function is equal to 1, and the results simplify: for the case $n=N=3$
see in~\cite{Nickel}, and for the case $n=N=4$ see in~\cite{OW,Wagner}. Other interesting
examples of using the geometrical approach can be found, e.g., in~\cite{other_geom}.

In this paper we will show that the natural way of splitting the basic simplex, 
as prescribed within the geometrical approach discussed above, leads to a reduction
of the effective number of independent variables in separate contributions obtained 
as a result of such splitting.

\section{Two-point function}

\begin{figure}[h]
\begin{minipage}{16pc}
\vspace*{10mm}
\includegraphics[width=16pc]{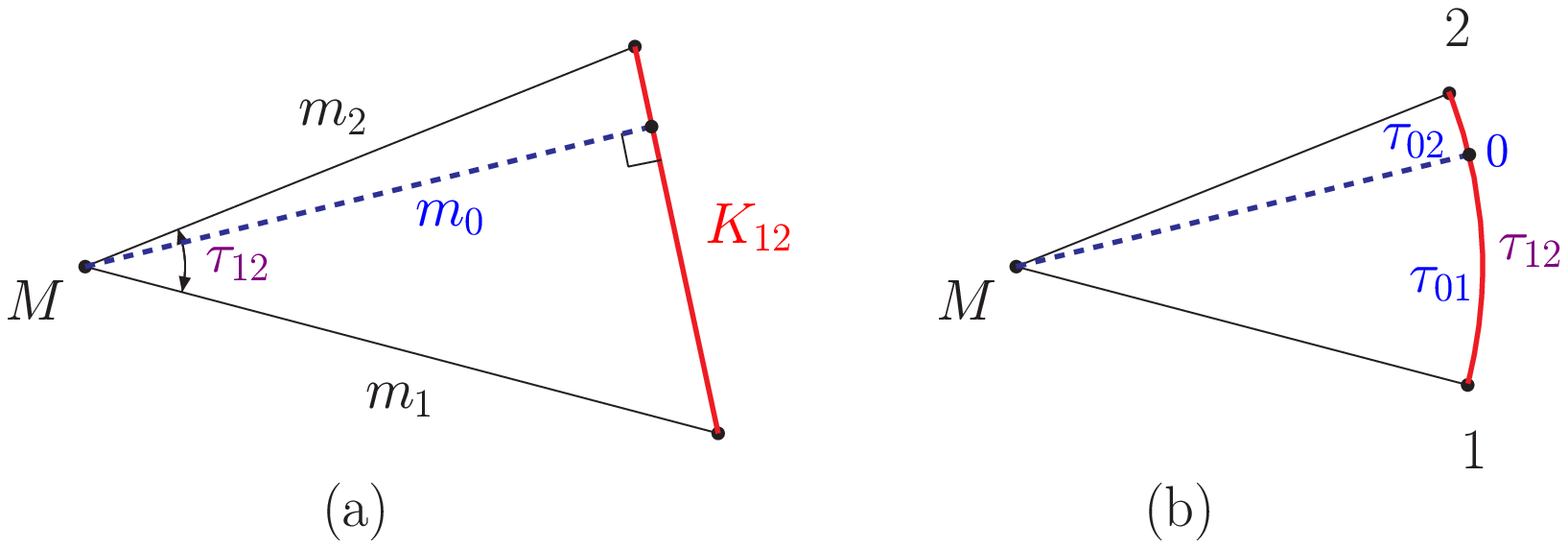}
\caption{\label{label}Two-point case: (a) the basic triangle and (b) the arc $\tau_{12}$.}
  \label{fig2pt1}
\end{minipage}\hspace{2pc}%
\begin{minipage}{19pc}
\vspace*{7mm}
\includegraphics[width=19pc]{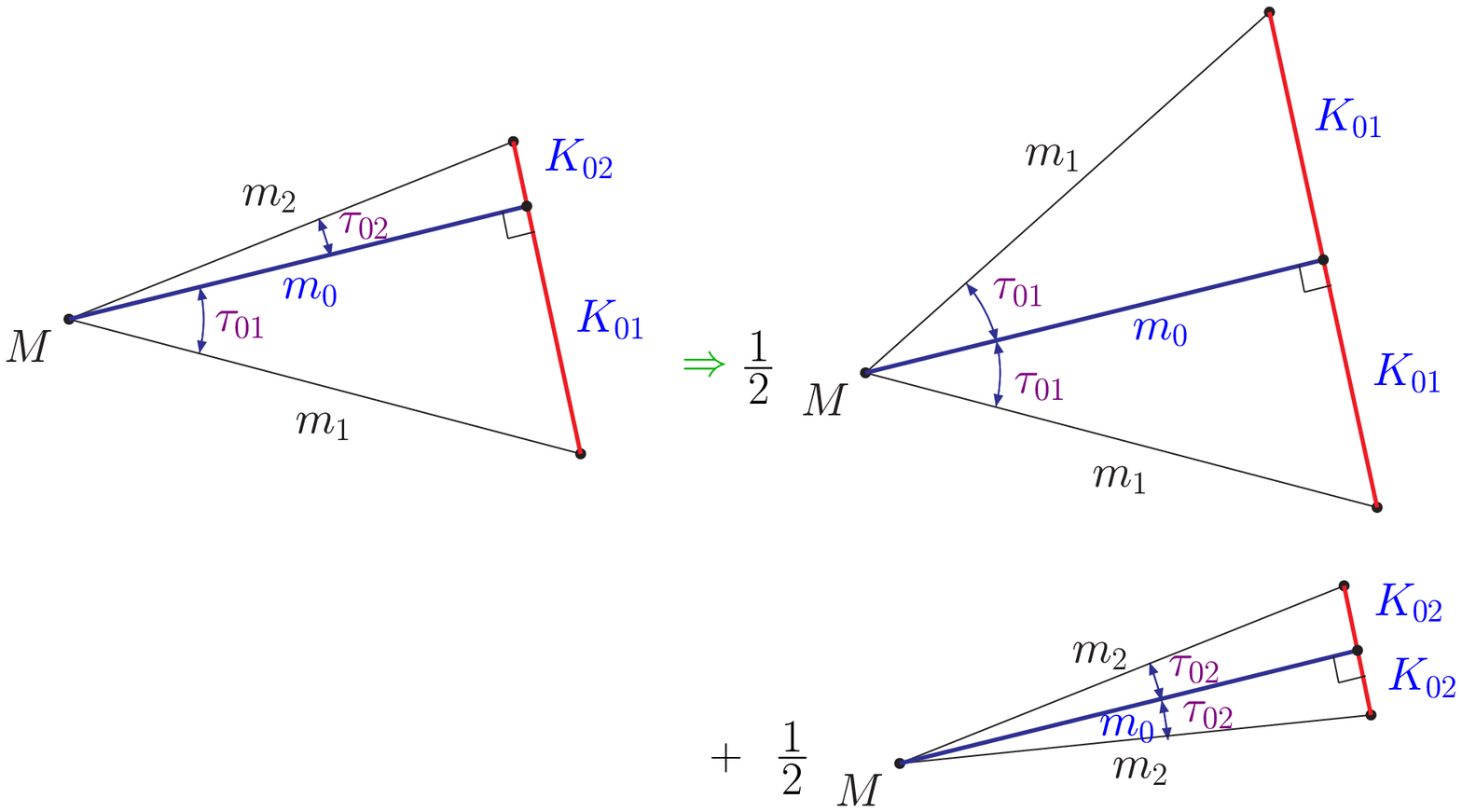}
\vspace*{-15mm}
\caption{\label{label}Two-point function: reduction to equal-mass integrals.}
  \label{fig2pt2}
\end{minipage} 
\end{figure}

For the two-point function, there is only one external momentum invariant $k_{12}^2$,
and the sides of the corresponding basic triangle are $m_1$, $m_2$ and $K_{12}\equiv\sqrt{k_{12}^2}$,
as shown in figure~\ref{fig2pt1}a. The angle $\tau_{12}$ between the sides $m_1$ and $m_2$
is defined through $\cos\tau_{12}\equiv c_{12}=(m_1^2+m_2^2-k_{12}^2)/(2m_1 m_2)$, and 
(in the spherical case) the integration goes over the arc $\tau_{12}$ of the unit circle,
as shown in figure~\ref{fig2pt1}b.

For splitting we use the height of the basic triangle, $m_0$, and obtain two triangles with 
the sides ($m_1$, $m_0$, $K_{01}\equiv\sqrt{k_{01}^2}$) and 
($m_2$, $m_0$, $K_{02}\equiv\sqrt{k_{02}^2}$),
respectively. Here $m_0=m_1 m_2 \sin{\tau_{12}}/\sqrt{k_{12}^2}$, 
$k_{01}^2 = (k_{12}^2+m_1^2-m_2^2)^2/(4 k_{12}^2)$ and
$k_{02}^2 = (k_{12}^2-m_1^2+m_2^2)^2/(4 k_{12}^2)$ (note that $k_{01}^2=m_1^2-m_0^2$ and
$k_{02}^2=m_2^2-m_0^2$).  Each of the resulting integrals
can be associated with a two-point function, and we arrive at the following decomposition:  
\begin{eqnarray}
J^{(2)}\left(n; 1,1| k_{12}^2; m_1, m_2 \right) &=&
\frac{1}{2k_{12}^2}
\left\{
(k_{12}^2+m_1^2-m_2^2)
J^{(2)}\left(n; 1,1| k_{01}^2; m_1, m_0 \right)
\right.
\nonumber \\ && \qquad
\left.
+ (k_{12}^2-m_1^2+m_2^2)
J^{(2)}\left(n; 1,1| k_{02}^2; m_2, m_0 \right)
\right\} \; .
\end{eqnarray}
This is an example of a functional relation between integrals
with different momenta and masses, similar to those described in~\cite{Tarasov2}.
Moreover, using the geometrical relation shown in figure~3, we can represent 
the right-hand side in terms of the equal-mass integrals:
\begin{eqnarray}
J^{(2)}\left(n; 1,1| k_{12}^2; m_1, m_2 \right) &=&
\frac{1}{4k_{12}^2}
\left\{
(k_{12}^2+m_1^2-m_2^2)
J^{(2)}\left(n; 1,1| 4 k_{01}^2; m_1, m_1 \right) 
\right.
\nonumber \\ && \qquad
\left.
+ (k_{12}^2-m_1^2+m_2^2)
J^{(2)}\left(n; 1,1| 4 k_{02}^2; m_2, m_2 \right)
\right\} \; .
\end{eqnarray}

Let us look at the number of variables. In the original integral 
$J^{(2)}\left(n; 1,1| k_{12}^2; m_1, m_2 \right)$ we have three 
independent variables: two masses and one momentum invariant 
(out of them we can construct two dimensionless variables).
In the integral $J^{(2)}\left(n; 1,1| k_{01}^2; m_1, m_0 \right)$
we have one extra condition on the variables, $k_{01}^2=m_1^2-m_0^2$,
so that we get two independent variables (i.e., 
one dimensionless variable). The same is valid for 
$J^{(2)}\left(n; 1,1| 4 k_{01}^2; m_1, m_1 \right)$, where 
the extra condition is due to two equal masses. Therefore,
the result for the two-point function in arbitrary dimension 
can be expressed in terms of a combination of functions of
a single dimensionless variable: indeed, we know that it can be presented
in terms of the Gauss hypergeometric function $_2F_1$
(see, e.g., in~\cite{DD-JMP,BDS}) whose $\varepsilon$-expansion
is known to any order~\cite{D-ep,DK1}. 

\section{Three-point function}

\begin{figure}[h]
\begin{minipage}{18pc}
\vspace*{10mm}
\includegraphics[width=18pc]{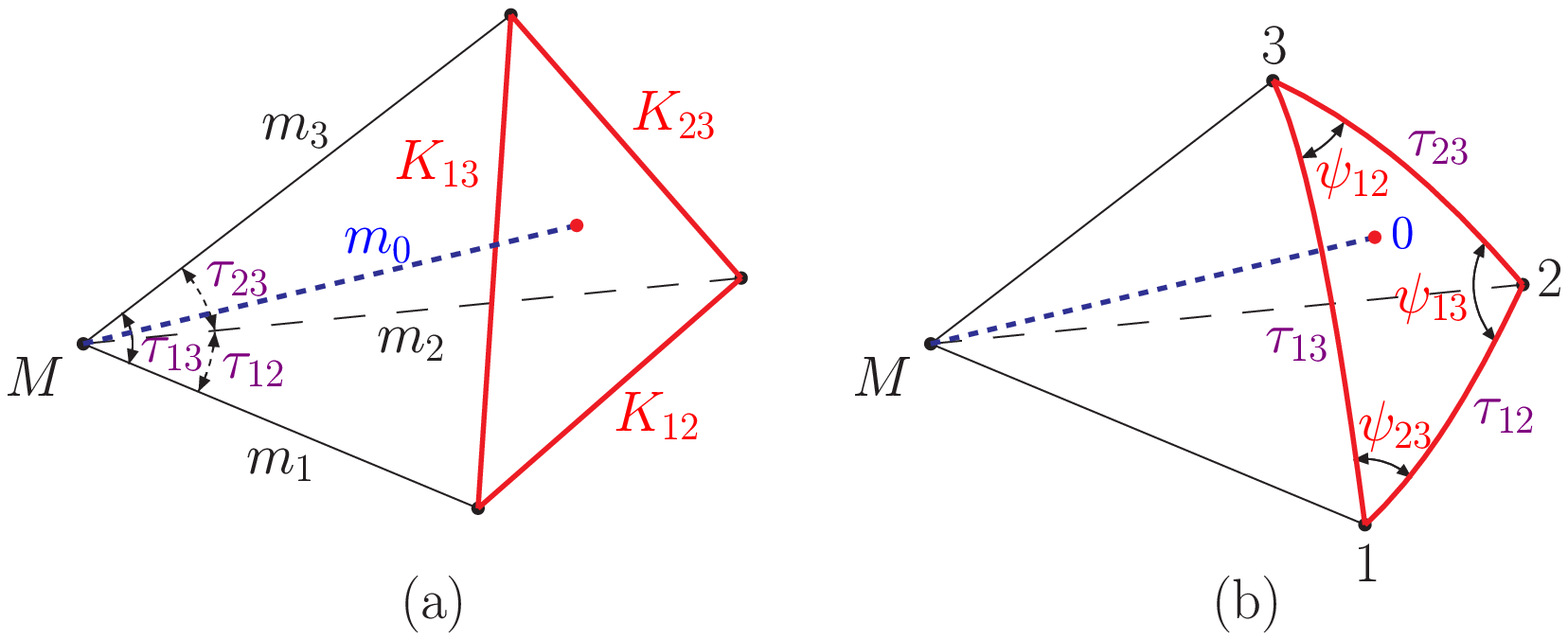}
\vspace*{-15mm}
\caption{\label{label}Three-point case: (a) the basic tetrahedron 
and (b) the solid angle.}
  \label{fig3pt1}
\end{minipage}\hspace{2pc}%
\begin{minipage}{18pc}
\vspace*{11mm}
\includegraphics[width=18pc]{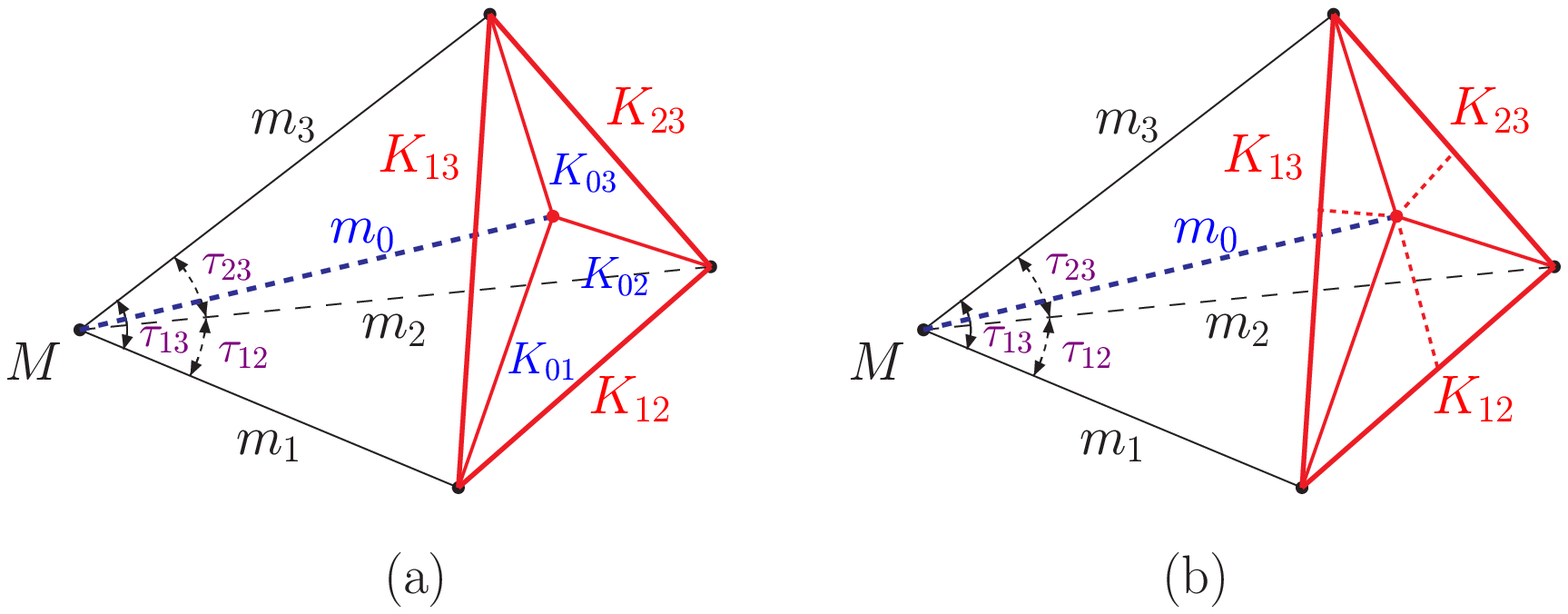}
\vspace*{-15mm}
\caption{\label{label}(a) Splitting the basic tetrahedron into three tetrahedra 
and (b) further splitting into six tetrahedra.}
  \label{fig3pt2}
\end{minipage} 
\end{figure}

For the three-point function, there are three external momentum invariants, $k_{12}^2$,
$k_{13}^2$ and $k_{23}^2$,
and the sides of the corresponding basic tetrahedron are $m_1$, $m_2$, $m_3$, 
$K_{12}\equiv\sqrt{k_{12}^2}$,
$K_{13}\equiv\sqrt{k_{13}^2}$ and $K_{23}\equiv\sqrt{k_{23}^2}$,
as shown in figure~\ref{fig3pt1}a. The angles $\tau_{12}$, $\tau_{13}$ and $\tau_{23}$
between the sides $m_1$, $m_2$ and $m_3$
are defined through $\cos\tau_{jl}\equiv c_{jl}=(m_j^2+m_l^2-k_{jl}^2)/(2m_j m_l)$, and 
(in the spherical case) the integration extends over the spherical triangle 123 of the unit sphere,
see in figure~\ref{fig3pt1}b.

For the splitting we use the height of the basic tetrahedron, $m_0$, and obtain three tetrahedra,
as shown in figure~5a. 
One of them has the sides $m_1$, $m_2$, $m_0$, $K_{12}\equiv\sqrt{k_{12}^2}$, 
$K_{01}\equiv\sqrt{k_{01}^2}$ and $K_{02}\equiv\sqrt{k_{02}^2}$,
and the sides for the others can be obtained by permutation of the indices.
Here $k_{01}^2 = m_1^2 - m_0^2$, $k_{02}^2 = m_2^2 - m_0^2$, $k_{03}^2 = m_3^2 - m_0^2$, and
$m_0=m_1 m_2 m_3 \sqrt{D^{(3)}/\Lambda^{(3)}}$, where $\Lambda^{(3)} = {\textstyle{\frac{1}{4}}}
\left[ 2 k_{12}^2 k_{13}^2 + 2 k_{13}^2 k_{23}^2 + 2 k_{23}^2 k_{12}^2
- (k_{12}^2)^2 - (k_{13}^2)^2 - (k_{23}^2)^2 \right]$, and $D^{(3)}=\det\|c_{jl}\|$ is 
the Gram determinant, see in~\cite{DD-JMP,D-NIMA} for more details.
Each of the resulting integrals
can be associated with a specific three-point function, and we arrive at the following decomposition:
\begin{eqnarray}
J^{(3)}\!\left( n; 1,1,1\big| k_{23}^2, k_{13}^2, k_{12}^2; m_1, m_2, m_3 \!\right)
&\!\!\!\!=\!\!\!& 
\frac{m_1^2 m_2^2 m_3^2}{\Lambda^{(3)}}
\Biggl\{
\frac{F_1^{(3)}}{m_1^2}
J^{(3)}\!\left( n; 1,1,1\big| k_{23}^2, k_{03}^2, k_{02}^2; m_0, m_2, m_3 \!\right)
\hspace*{-2mm}
\nonumber \\ && 
+ \frac{F_2^{(3)}}{m_2^2}
J^{(3)}\!\left( n; 1,1,1\big| k_{03}^2, k_{13}^2, k_{01}^2; m_1, m_0, m_3 \right)
\nonumber \\ && 
+ \frac{F_3^{(3)}}{m_3^2}
J^{(3)}\!\left( n; 1,1,1\big| k_{02}^2, k_{01}^2, k_{12}^2; m_1, m_2, m_0 \right)
\Biggr\} \; ,
\end{eqnarray}
with
\begin{equation}
F_3^{(3)}
= \frac{1}{4m_1^2m_2^2}
\Bigl[
k_{12}^2 \left( k_{13}^2 \!+\! k_{23}^2 \!-\! k_{12}^2 
\!+\! m_1^2 \!+\! m_2^2 \!-\! 2 m_3^2\right)
- (m_1^2-m_2^2) \left( k_{13}^2 - k_{23}^2 \right)
\Bigr], 
\end{equation}
etc., so that 
$\sum_{i=1}^3 (F_i^{(3)}/m_i^2) = \Lambda^{(3)}/(m_1^2 m_2^2 m_3^2)$.

By dropping perpendiculars onto the sides $K_{12}\equiv\sqrt{k_{12}^2}$, etc., 
each of the resulting tetrahedra can be split into two, so that in
total we get six ``birectangular" tetrahedra, as shown in figure~5b.
Furthermore, for each of them we can use the geometrical relation 
similar to one shown in figure~3, reducing them to the integrals with
two equal masses:
\begin{eqnarray}
&& \hspace*{-15mm}
J^{(3)}\left( n; 1,1,1 \big| k_{02}^2, k_{01}^2, 
k_{12}^2; m_1, m_2, m_0\right)
\nonumber \\  
&\!\!=\!\!&
\frac{1}{2 k_{12}^2}
\Biggl\{
(k_{12}^2 + m_1^2 - m_2^2)
J^{(3)}\left( n; 1,1,1 \big| k_{01}^2, k_{01}^2, 
\frac{(k_{12}^2+m_1^2-m_2^2)^2}{k_{12}^2}; m_1, m_1, m_0\right)
\nonumber \\ && \qquad
+ (k_{12}^2 - m_1^2 + m_2^2)
J^{(3)}\left( n; 1,1,1 \big| k_{02}^2, k_{02}^2, 
\frac{(k_{12}^2-m_1^2+m_2^2)^2}{k_{12}^2}; m_2, m_2, m_0\right)
\Biggr\} \; .
\end{eqnarray}

Let us analyze the number of variables. In the integral 
$J^{(3)}\left( n; 1,1,1\big| k_{23}^2, k_{13}^2, k_{12}^2; m_1, m_2, m_3 \right)$ 
we have six independent variables: three masses and three momentum invariants 
(out of them we can construct five dimensionless variables).
In $J^{(3)}\left( n; 1,1,1 \big| k_{02}^2, k_{01}^2, 
k_{12}^2; m_1, m_2, m_0\right)$
we have two extra conditions on the variables, $k_{01}^2=m_1^2-m_0^2$ 
and $k_{02}^2=m_2^2-m_0^2$,
so that we get four independent variables (i.e., 
three dimensionless variables). 
For the integral  
$J^{(3)}\left( n; 1,1,1 \big| k_{01}^2, k_{01}^2, 
(k_{12}^2+m_1^2-m_2^2)^2/k_{12}^2; m_1, m_1, m_0\right)$
we have three relations, with 
an additional condition due to the two equal masses. Therefore,
the result for the three-point function in arbitrary dimension 
should be expressible in terms of a combination of functions of
two dimensionless variables: indeed, we know that it can be presented
in terms of the Appell hypergeometric function $F_1$
(see, e.g., in~\cite{D-NIMA,Tarasov-NPBPS,FJT}).

\section{Four-point function}

For the four-point function, there are six external momentum invariants. 
Out of them, $k_{12}^2$, $k_{23}^2$, $k_{34}^2$ and $k_{14}^2$ are the squared
momenta of the external legs, whilst $k_{13}^2$ and $k_{24}^2$ correspond
to the Mandelstam variables $s$ and $t$. The sides of the corresponding basic 
four-dimensional simplex are $m_1$, $m_2$, $m_3$, $m_4$, and six additional
sides $K_{jl}\equiv\sqrt{k_{jl}^2}$,
as shown in figure~6a. The six angles $\tau_{jl}$
between the corresponding sides $m_j$ and $m_l$
are defined through $\cos\tau_{jl}\equiv c_{jl}=(m_j^2+m_l^2-k_{jl}^2)/(2m_j m_l)$, and 
(in the spherical case) the integration extends over the spherical tetrahedron 1234 
of the unit hypersphere, as shown in figure~6b (for the hyperbolic case
one can use analytic continuation).

\begin{figure}[h]
\vspace*{10mm}
\begin{center}
\includegraphics[width=30pc]{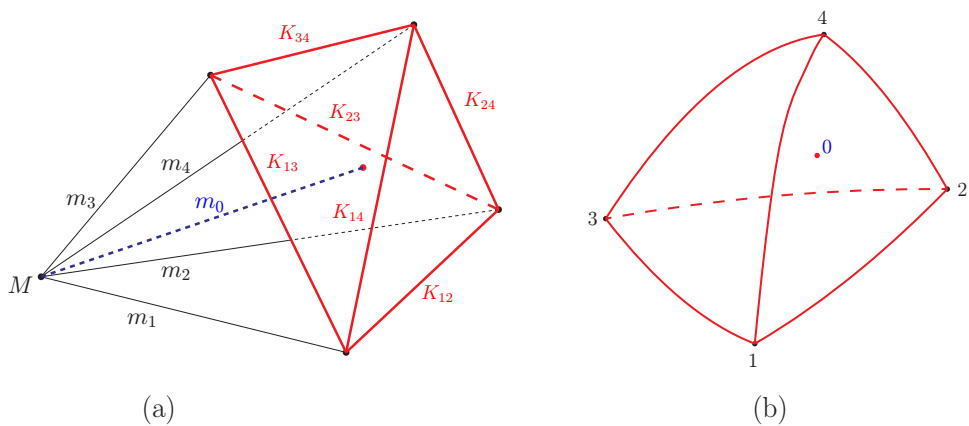}
\end{center}
  \label{fig4pt1}
\vspace*{-15mm}
\caption{\label{label}Four-point case: (a) the basic simplex 
and (b) the spherical tetrahedron.}
\end{figure}

For splitting we use the height of the basic simplex, $m_0$, and obtain four simplices,
as shown in figure~7a. 
One of them has the sides $m_1$, $m_2$, $m_3$, $m_0$, $K_{12}\equiv\sqrt{k_{12}^2}$, 
$K_{13}\equiv\sqrt{k_{13}^2}$, $K_{23}\equiv\sqrt{k_{23}^2}$, 
$K_{01}\equiv\sqrt{k_{01}^2}$, $K_{02}\equiv\sqrt{k_{02}^2}$ and $K_{03}\equiv\sqrt{k_{03}^2}$,
and the sides of the others can be obtained by permutation of the indices.
As before, $k_{0i}^2 = m_i^2 - m_0^2$ ($i=1,2,3,4$), whereas
$m_0=m_1 m_2 m_3 m_4\sqrt{D^{(4)}/\Lambda^{(4)}}$, where $D^{(4)}=\det\|c_{jl}\|$ and 
$\Lambda^{(4)} = \det\|(k_{j4}\cdot k_{l4})\|$, 
see in~\cite{DD-JMP} for more details.
Each of the four resulting integrals
can be associated with a certain four-point function.
At the next step, in each of the four tetrahedra (drawn in red) we drop the perpendiculars 
onto the triangle sides, as shown in figure~7b, splitting each of them into three, 
and then dividing each of the resulting
tetrahedra into two, by dropping perpendiculars onto the $\sqrt{k_{jl}^2}$ sides, 
as shown in figure~7c.
As a result of this splitting, we get 24 simplices.
The corresponding steps of splitting the spherical tetrahedron are shown in figure~8.

\begin{figure}[h]
\vspace*{10mm}
\includegraphics[width=30pc]{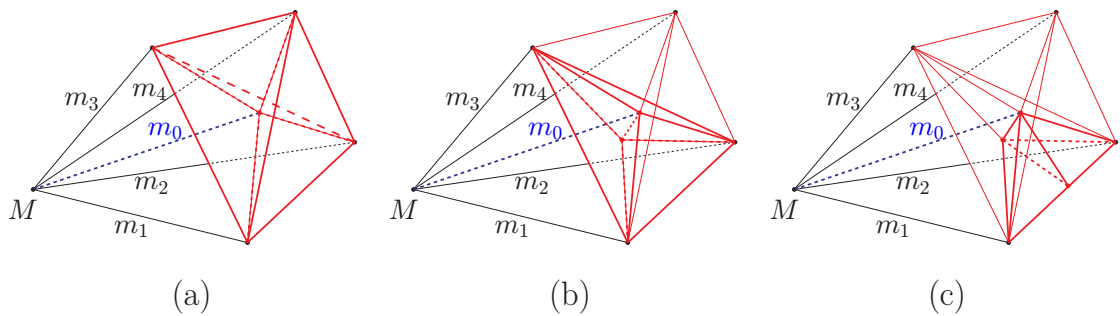}
\vspace*{-3mm}
\caption{\label{label}Four-point case: splitting the basic four-dimensional simplex.}
\end{figure}

\begin{figure}[h]
\begin{center}
\vspace*{-5mm}
\includegraphics[width=30pc]{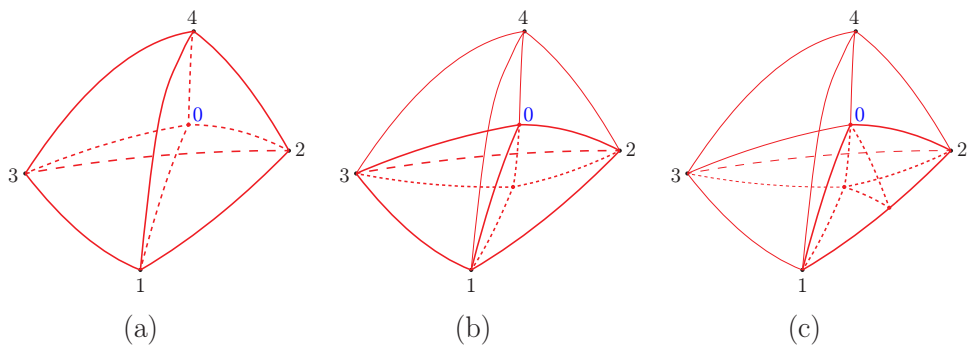}
\end{center}
\vspace*{-5mm}
\caption{\label{label}Four-point case: splitting the spherical tetrahedron.}
\end{figure}

Let us look at the number of variables. In the integral 
$J^{(4)}\left( n; 1,1,1,1\big| \{k_{jl}^2\}; \{m_i\} \right)$ 
we have ten independent variables: four masses and six momentum invariants 
(out of them we can construct nine dimensionless variables).
After the first step (figure~7a)
we have three extra conditions on the variables, $k_{01}^2=m_1^2-m_0^2$, $k_{02}^2=m_2^2-m_0^2$ 
and $k_{03}^2=m_3^2-m_0^2$,
so that we get seven independent variables (i.e., 
six dimensionless variables). 
After the second step (figure~7b), we get two extra conditions due to the right triangles,
and after the third step (figure~7c) we get one more condition.
As a result, for each of the 24 resulting four-point functions we have six relations,
so that we end up with four independent variables
(i.e., three dimensionless variables). 
Therefore, the result for the four-point function in arbitrary dimension 
should be expressible in terms of a combination of functions of
three dimensionless variables, such as, e.g., Lauricella functions and 
their generalizations (see, e.g., in~\cite{FJT,BKM}).

\section{General remarks and conclusions}

Using a geometrical approach,
we can relate the one-loop $N$-point Feynman diagrams to certain 
volume integrals in non-Euclidean geometry. 
Geometrical splitting provides a straightforward way of reducing
general integrals to those with lesser number of independent
variables. In this way, we can predict the set and the number of these variables
in the resulting integrals. Furthermore, it allows us to derive functional relations
between integrals with different momenta and masses. 

Numbers of dimensionless variables in separate contributions for $N$-point
diagrams, before and after the splitting, are summarized in the table.
\begin{table}[h]
\begin{center}
\caption{Number of variables before and after the splitting}
\begin{tabular}{|c|c|c|c|}
\hline
$\;$ & {\footnotesize  total \# of } & {\footnotesize \# of splitting} & {\footnotesize reduced \# } \\ 
$\;$ & {\footnotesize dimensionless variables} & {\footnotesize pieces} & {\footnotesize of variables} \\ 
\hline
$N=2$ & $3-1=2$ & 2 & 1 \\
\hline
$N=3$ & $6-1=5$ & 6 & 2 \\
\hline
$N=4$ & $10-1=9$ & 24 & 3 \\
\hline
arbitrary $N$ & $\tfrac{1}{2}(N-1)(N+2)$ & $N!$ & $N-1~~(?)$ \\
\hline
\end{tabular}
\end{center}
\vspace*{-5mm}
\end{table}

\section*{Acknowledgements}
I am thankful to R~Delbourgo and M~Yu~Kalmykov with whom I started to work on this subject. I am 
grateful to the organizers of ACAT-2016 and the University of B\'{i}o-B\'{i}o for their support and hospitality.

\end{document}